\documentstyle[prl,aps,psfig,multicol]{revtex}
\begin{document}
\draft

\title{Successive opening of the Fermi surface in doped $N$-leg Hubbard ladders}
\author{Urs Ledermann, Karyn Le Hur, and T.M. Rice}
\address{Theoretische Physik, Eidgen\"{o}ssische Technische Hochschule,
CH-8093 Z\"{u}rich, Switzerland}

\date{\today}
\maketitle

\begin{abstract}
We study the effect of doping away from half-filling in weakly (but finitely)
interacting $N$-leg Hubbard ladders using renormalization group and bosonization
techniques. For a small on-site repulsion $U$, the $N$-leg Hubbard ladders are
equivalent to a $N$-band model, where at half-filling the Fermi velocities are
$v_{1}=v_{N}<v_{2}=v_{N-1}<\ldots$. We then obtain a hierarchy of energy-scales,
where the band pairs $(j,N+1-j)$ are successively frozen out. The low-energy
Hamiltonian is then the sum of $N/2$ (or $(N-1)/2$ for $N$ odd) two-leg ladder
Hamiltonians without gapless excitations (plus a single chain for $N$ odd with
one gapless spin mode) --- similar to the $N$-leg Heisenberg spin-ladders. The
energy-scales lead to a hierarchy of gaps. Upon doping away from half-filling,
the holes enter first the band(s) with the smallest gap: For odd $N$, the holes
enter first the nonbonding band $(N+1)/2$ and the phase is a Luttinger liquid,
while for even $N$, the holes enter first the band pair $(N/2,N/2+1)$ and the
phase is a Luther-Emery liquid, similar to numerical treatments of the $t$-$J$
model, i.e., at and close to half-filling, the phases of the Hubbard ladders for
small and large $U$ are the same. For increasing doping, hole-pairs subsequently
enter at critical dopings the other band pairs $(j,N+1-j)$ (accompanied by a
diverging compressibility): The Fermi surface is successively opened by doping,
starting near the wave vector $(\pi/2,\pi/2)$. Explicit calculations are given
for the cases $N=3,4$.
\end{abstract}


\begin{multicols}{2}

\narrowtext

\section{Introduction}
Interest in the field of ladder systems grew rapidly, when the possibility of superconductivity
in doped two-leg ladder materials was proposed \cite{bib:DRS,bib:RGS}. Experimentally, such a
behavior has been observed in a two-leg ladder material under high pressure~\cite{bib:UNA}.
The $N$-leg ladders have then been studied both as a first step towards the two-dimensional (2D)
case \cite{bib:WStJ} and because of their unusual physical properties. For example, spin-ladders
exhibit an odd-even effect~\cite{bib:DagRice}; while even-leg ladders have a spin-gap, odd-leg
ladders have one gapless spin mode. Similarly, the behavior upon doping is very different. The
two-leg ladder is a Luther-Emery liquid, but the lightly doped three-leg ladder consists of a
Luttinger liquid (LL) plus an insulating spin liquid (ISL)~\cite{bib:ThreeLegLD,bib:WSThreeLeg}.
The latter is due to a truncation of the Fermi surface (FS) in two channels. Such a truncation
is argued to take place also in the 2D case, where upon doping some parts of the FS
become conducting and other parts remain an ISL~\cite{bib:FRS}.

Analytic works for (doped) ladders are usually based on the Hubbard model and restricted
to a weak on-site repulsion $U$ (for large $U$, see Ref.~\cite{bib:ShTs}). Insulating
behavior at half-filling is then due to umklapp processes. The two-leg ladder has been
investigated by many authors over the past few
years~\cite{bib:Fabrizio,bib:LBFso8,bib:KLLS,bib:Schulzxxz}. We note that in the lightly
doped two-leg ladder paired holes form a dilute gas of hard-core bosons, which are equivalent
to spinless fermions~\cite{bib:LBFso8,bib:Schulzxxz}. Away from half-filling, the phases of
the weak interaction limit $U\rightarrow 0$ of the three- and four-leg ladder have been
derived in Refs.~\cite{bib:Arrigoni,bib:LBF} (for possible superconductivity,
see Ref.~\cite{bib:KKA}). For the three-leg ladder, a C2S1 phase has been
obtained upon doping (a phase with $n$ gapless charge- and $m$ gapless spin modes is denoted
by C$n$S$m$). This contrasts numerical findings for the (strongly interacting) $t$-$J$ model,
where at low dopings the phase is C1S1 \cite{bib:ThreeLegLD,bib:WSThreeLeg}. An idea to
explain this discrepancy has been given in a related context, where a one-dimensional electron
gas (1DEG) was coupled to an insulating ``environment'' (another 1DEG)~\cite{bib:EKZ}.

In this paper, we study the doping $\delta$ away from half-filling of $N$-leg
Hubbard ladders in the case of weak but \emph{finite} repulsive interactions,
$0<U\ll t_{\perp},t$ ($t$ and $t_{\perp}$ denote the hopping matrix elements along- and
between the chains). Previous works have taken the limit $U\rightarrow 0$ first, keeping
the doping $\delta$ finite~\cite{bib:Arrigoni,bib:LBF}. Here, we treat the opposite limit,
i.e., we keep $U$ finite and then investigate the lightly doped case $\delta\rightarrow 0$,
where the umklapp interactions present at half-filling cannot be neglected. It is then
advantageous to take the half-filled $N$-leg ladders as a starting point.

For a small $U$, the Hubbard-model reduces to a weakly interacting $N$-band model. After
linearization around the Fermi points, each band is characterized by a Fermi velocity
$v_{j}$; at half-filling $v_{1}=v_{N}<v_{2}=v_{N-1}<\ldots$. The low-energy
physics is then derived by a combination of the renormalization group (RG)
\cite{bib:Shankar} and bosonization methods \cite{bib:Haldane,bib:SchulzRev}.
Integrating the RG equations (RGEs) at half-filling, we find a
\emph{decoupling into band pairs} $(j,N+1-j)$ and a hierarchy of energy-scales
$T_{j}\sim te^{-\alpha v_{j}/U}$ ($\alpha$ is a constant of the order of 1),
where the band pairs scale towards a two-leg ladder fixed
point and become frozen out (this behavior was not
noted in previous works~\cite{bib:Arrigoni,bib:LBF}; in particular, Ref.~\cite{bib:LBF}
concentrated on doping levels (well) away from half-filling).
For $N$ even, all excitations are then gapped. For odd $N$, the remaining band behaves like a
single chain such that the charge degrees of freedom acquire a gap while the
spin-excitations stay gapless. We therefore recover the odd-even effect that
is present in Heisenberg spin-ladders.

The gaps for the band pairs are given by $\Delta_{j}\sim T_{j}$ resulting in a
hierarchy of gaps. Upon doping, the holes enter first the band(s) with the
smallest gap. For odd $N$, the gap of the single chain is the smallest one and at
lowest doping, the phase is a LL (C1S1). For even $N$, the spin degrees of freedom
remain gapped upon doping and the phase is a Luther-Emery liquid (C1S0). These
phases are the same as obtained in (numerical) treatments of the strongly interacting
$t$-$J$ model~\cite{bib:WStJ,bib:ThreeLegLD,bib:WSThreeLeg,bib:WSFourLeg,bib:SchAnalyt}, i.e., at
and close to half-filling, the phases for small and large $U$ are the same. For
increasing doping, paired holes enter at critical dopings, where each time the
compressibility diverges, subsequently the other band pairs $(j,N+1-j)$. In other
words, the FS is \emph{successively} opened, first near the wave vector
$(\pi/2,\pi/2)$, spreading towards the wave vectors $(\pi,0)$ and $(0,\pi)$ for
increasing doping, see Figs.~\ref{f:FermiS3Leg} and~\ref{f:FermiS4Leg}. Explicit
calculations will be given for the cases~$N=3,4$.

In Sec. II, the $N$-leg Hubbard model is introduced and the phases at half-filling are
derived. The main result of this section is a low-energy Hamiltonian allowing us to
study the effect of doping away from half-filling: In Sec.~III, including the doping
as a perturbation of the  half-filled Hamiltonian, we treat first the case when only
one band (pair) is doped and then the case when many band pairs become doped.

\section{Half filled $N$-leg Hubbard ladders}

\subsection{The Hamiltonian}
The $N$-leg Hubbard model is given by $H=H_{0}+H_{\rm Int}$, where the kinetic energy is
\begin{eqnarray}
  H_{0}&=&-t\sum_{x,j,s} d_{js}^{\dagger}(x+1)d_{js}(x)
    +{\mathrm H.c.}
  \nonumber\\
  &&-t_{\perp}\sum_{x,j,s}d_{j+1s}^{\dagger}(x)d_{js}(x)
    +{\mathrm H.c.}
\end{eqnarray}
and the interaction term is
\begin{equation}
  H_{\rm Int}=U\sum_{j,x}d_{j\uparrow}^{\dagger}(x)d_{j\uparrow}(x)
    d_{j\downarrow}^{\dagger}(x)d_{j\downarrow}(x).
\end{equation}
The hopping matrix elements along- and perpendicular to the legs are
denoted by $t$ and $t_{\perp}$ respectively and $U>0$ is the on-site repulsion.
For weak interactions, $U\ll t_{\perp},t$, it is advantageous first to 
diagonalize $H_{0}$,
\begin{equation}
  H_{0}=\sum_{j,s}\int dk\epsilon_{j}(k)\Psi_{js}^{\dagger}(k)
    \Psi_{js}(k),
\end{equation}
where $\Psi_{js}^{\dagger}$ and $\Psi_{js}$ are the creation-
and annihilation operators for the band~$j$.

\begin{figure}[t]
  \centerline{
    \psfig{file=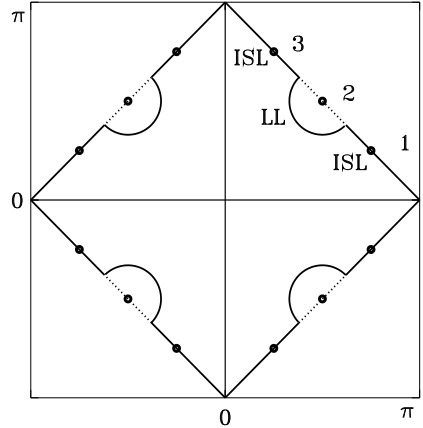,width=6.3cm,height=6.1cm}}
  \vspace{1mm}
  \caption{The bands of the three-leg ladder correspond to three different points on a 2D FS
  (denoted by 1, 2, and 3). The square is the umklapp surface, which is the FS at half-filling
  (for $t=t_{\perp}$). The half-circles indicate the opening of the FS, when the band two is
  doped. At (close to) the wave vector $(\pi/2,\pi/2)$, the phase is then a LL, while at
  $(3\pi/4,\pi/4)$ and $(\pi/4,3\pi/4)$, the phase is an ISL.}
  \label{f:FermiS3Leg}
\end{figure}

\noindent
For open boundary conditions perpendicular to the legs, the dispersion relations $\epsilon_{j}$
are given by ($j=1,\ldots,N$)
\begin{equation}
  \epsilon_{j}(k)=-2t\cos(k)-2t_{\perp}\cos\left[\pi j/(N+1)\right].
  \label{eq:DispRelN}
\end{equation}
The Fermi momenta in each band $k_{Fj}$ are determined by the chemical
potential $\mu=\epsilon_{j}(k_{Fj})$ and the filling $n=2(\pi N)^{-1}\sum k_{Fj}$.
Since we are only interested in the low-energy physics, we linearize
the dispersion relation around the Fermi momenta, resulting in Fermi velocities
$v_{j}=2t\sin(k_{Fj})$. We introduce operators $\Psi_{R/Ljs}$ for right- and left
movers at the Fermi level in each band.

At half-filling $n=1$, $\mu=0$ and the velocities are given by
\begin{equation}
  v_{j}=v_{\bar{\jmath}}=2\sqrt{t^{2}-\left\{t_{\perp}\cos[\pi j/(N+1)]\right\}^{2}},
  \label{eq:HFVel}
\end{equation}
where $\bar{\jmath}=N+1-j$. Note that
\begin{equation}
  v_{1}=v_{N}<v_{2}=v_{N-1}<\ldots.
  \label{eq:HFVelNEQ}
\end{equation}
These particular values of the Fermi velocities will lead to a hierarchy of
energy scales (see below).

For \emph{generic} Fermi momenta, the interactions are forward-
($f$) and Cooper- ($c$) scattering within a band or between two different bands.
Following Ref.~\cite{bib:LBF}, the interacting part of the Hamiltonian can then be written as

\begin{figure}[b]
  \centerline{
    \psfig{file=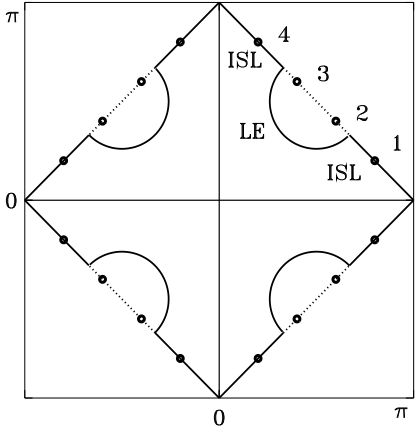,width=6.3cm,height=6.1cm}}
  \vspace{1mm}
  \caption{The bands of the four-leg ladder correspond to four different points on a 2D FS
  (denoted by 1, 2, 3, and 4). The square is the umklapp surface, which is the FS at half-filling
  (for $t=t_{\perp}$). The half-circles indicate the opening of the FS, when the bands two and
  three are doped, forming a Luther-Emery (LE) liquid, where the pairing takes place between
  quasi-particles at (close to) the wave vectors $(2\pi/5,3\pi/5)$ and $(3\pi/5,2\pi/5)$. At
  $(4\pi/5,\pi/5)$ and $(\pi/5,4\pi/5)$, the phase is still an ISL. Similarly as for the two-leg
  ladder, the order-parameter has a different sign in the Fermi points~2 and~3.}
  \label{f:FermiS4Leg}
\end{figure}
\begin{eqnarray}
  H_{\rm Int}&=&\sum_{i\neq j}\int dx\left(f_{ij}^{\rho}J_{Rii}J_{Ljj}
    -f_{ij}^{\sigma}{\mathbf J}_{Rii}\cdot{\mathbf J}_{Ljj}\right.
  \nonumber\\
  &&\left.+c_{ij}^{\rho}J_{Rij}J_{Lij}
    -c_{ij}^{\sigma}{\mathbf J}_{Rij}\cdot{\mathbf J}_{Lij}\right)
  \nonumber\\
    &&+\sum_{j}\int dx\left(c_{jj}^{\rho}J_{Rjj}J_{Ljj}
    -c_{jj}^{\sigma}{\mathbf J}_{Rjj}\cdot{\mathbf J}_{Ljj}\right),
  \label{eq:BInt}
\end{eqnarray}
where the U(1) and SU(2) currents are defined as
\begin{equation}
  J_{hij}=\sum_{s}\Psi_{his}^{\dagger}\Psi_{hjs}
  \label{eq:ChargeCurrents}
\end{equation}
and
\begin{equation}
  J_{hij}^{k}=\frac{1}{2}\sum_{s,s^{\prime}}\Psi_{his}^{\dagger}
    \tau_{ss^{\prime}}^{k}\Psi_{hjs^{\prime}},
  \label{eq:SpinCurrents}
\end{equation}
where again $h=R/L$ and the $\tau^{k}$ are the Pauli matrices ($k=x,y,z$). Due to symmetry,
$f_{ij}^{\rho,\sigma}=f_{ji}^{\rho,\sigma}$ and $c_{ij}^{\rho,\sigma}=c_{ji}^{\rho,\sigma}$.

In the half-filled case we must generalize the interacting Hamiltonian~(\ref{eq:BInt}). The
Fermi momenta of the bands $j$ and $\bar{\jmath}$ add up to $\pi$, $k_{Fj}+k_{F\bar{\jmath}}=\pi$, allowing
umklapp interactions to take place between these two bands (for $u_{j\bar{\jmath}\bar{\jmath} j}^{\rho,\sigma}$
we differ in notation from Ref.~\cite{bib:LBFso8} by a factor~2),
\begin{eqnarray}
  u_{j\bar{\jmath}\bar{\jmath} j}^{\rho}\left(I_{Rj\bar{\jmath}}^{\dagger}I_{L\bar{\jmath} j}+{\mathrm H.c.}\right)
  -u_{j\bar{\jmath}\bar{\jmath} j}^{\sigma}\left({\mathbf I}_{Rj\bar{\jmath}}^{\dagger}\cdot{\mathbf I}_{L\bar{\jmath} j}+{\mathrm H.c.}\right)
  \nonumber\\
    +u_{jj\bar{\jmath}\bar{\jmath}}^{\rho}\left(I_{Rjj}^{\dagger}I_{L\bar{\jmath}\bar{\jmath}}+I_{R\bar{\jmath}\bar{\jmath}}^{\dagger}I_{Ljj}
    +{\mathrm H.c.}\right).
  \label{eq:TwoBdUk}
\end{eqnarray}
We have used the definitions
\begin{equation}
  I_{hij}=\sum_{s,s^{\prime}}\Psi_{his}\epsilon_{ss^{\prime}}\Psi_{hjs^{\prime}}
\end{equation}
and
\begin{equation}
  I_{hij}^{k}=\frac{1}{2}\sum_{s,s^{\prime}}
    \Psi_{his}\left(\epsilon\tau^{k}\right)_{ss^{\prime}}\Psi_{hjs^{\prime}},
\end{equation}
where $\epsilon=-i\tau^{y}$.
For odd $N$, there exists a single-band umklapp term for the band $r=(N+1)/2$
\begin{equation}
  u_{rr}^{\rho}\left(I_{Rrr}^{\dagger}I_{Lrr}+{\mathrm H.c.}\right).
\end{equation}
In addition to these processes, there exist various interactions involving four different
bands or three different bands (only for $N$ odd). For the half-filled three-leg ladder,
three different three-band umklapp processes
\begin{eqnarray}
  u_{1223}^{\rho}\left(I_{R12}^{\dagger}I_{L23}+I_{R32}^{\dagger}I_{L21}+{\mathrm H.c.}\right)
  \nonumber\\
  +u_{2213}^{\rho}\left(I_{R22}^{\dagger}I_{L13}+I_{R13}^{\dagger}I_{L22}+{\mathrm H.c.}\right)
  \nonumber\\
  -u_{1223}^{\sigma}\left({\mathbf I}_{R12}^{\dagger}\cdot{\mathbf I}_{L23}+
    {\mathbf I}_{R32}^{\dagger}\cdot{\mathbf I}_{L21}+{\mathrm H.c.}\right)
\end{eqnarray}
and two different three-band non-umklapp processes
\begin{eqnarray}
  c_{1223}^{\rho}\left(J_{R12}J_{L23}+J_{R32}J_{L21}+{\mathrm H.c.}\right)
  \nonumber\\
  -c_{1223}^{\sigma}\left({\mathbf J}_{R12}\cdot{\mathbf J}_{L23}
   +{\mathbf J}_{R32}\cdot{\mathbf J}_{L21}+{\mathrm H.c.}\right)
\end{eqnarray}
take place and have to be added to Eq.~(\ref{eq:BInt}).

In the four-leg case, the four-band umklapp interactions read
\begin{eqnarray}
  u_{1234}^{\rho}\left(I_{R12}^{\dagger}I_{L34}+I_{R34}^{\dagger}I_{L12}+{\mathrm H.c.}\right)
  +u_{1324}^{\rho}\left(2\leftrightarrow 3\right)
  \nonumber\\
  +u_{1423}^{\rho}\left(I_{R14}^{\dagger}I_{L23}+I_{R23}^{\dagger}I_{L14}+{\mathrm H.c.}\right)
  \nonumber\\
  -u_{1234}^{\sigma}\left({\mathbf I}_{R12}^{\dagger}\cdot{\mathbf I}_{L34}+
    {\mathbf I}_{R34}^{\dagger}\cdot{\mathbf I}_{L12}+{\mathrm H.c.}\right)
  \nonumber\\
  -u_{1324}^{\sigma}\left(2\leftrightarrow 3\right)
\end{eqnarray}
and the four-band non-umklapp interactions take the form
\begin{eqnarray}
  c_{1234}^{\rho}\left(J_{R12}J_{L34}+J_{R43}J_{L21}+{\mathrm H.c.}\right)
  +c_{1324}^{\rho}\left(2\leftrightarrow 3\right)
  \nonumber\\
  -c_{1234}^{\sigma}\left({\mathbf J}_{R12}\cdot{\mathbf J}_{L34}
   +{\mathbf J}_{R43}\cdot{\mathbf J}_{L21}+{\mathrm H.c.}\right)
  \nonumber\\
  -c_{1324}^{\sigma}\left(2\leftrightarrow 3\right).
\end{eqnarray}
In our case $v_{1}=v_{4}$ and $v_{2}=v_{3}$ implying $c_{1234}^{\rho,\sigma}=c_{1324}^{\rho,\sigma}$,
$u_{1234}^{\rho,\sigma}=u_{1324}^{\rho,\sigma}$. The generalization to $N>4$ is straightforward
such that we do not reproduce it here.

The (initial) values of the couplings are of the order of the on-site repulsion $U$
(for $N=3$, see Appendix~A).

\subsection{Low-energy phases}
Using the RG method (and bosonization), we derive the low-energy phases at
half-filling and a Hamiltonian allowing us to study the effect of doping, i.e.,
doping is then included as a (small) perturbation. We have generalized the
results given in Refs.~\cite{bib:LBFso8,bib:LBF} in order to treat ladders with
$N>2$ legs at half-filling~\cite{bib:ULup}. For weak interactions, $U\ll t$, it
is sufficient to take the RGEs in one-loop order, i.e., quadratic in the
coupling constants.

Integrating the RGEs, we find, that the Fermi velocities $v_{j}$ [see
Eqs.~(\ref{eq:HFVel}) and (\ref{eq:HFVelNEQ})] lead to a hierarchy of
energy-scales by $T_{j}\sim te^{-\alpha v_{j}/U}$, where the band pairs $(j,\bar{\jmath})$
become successively frozen out. This exponential dependence on $v_{j}/U$ is a consequence of the
logarithmic derivatives of the couplings with respect to the energy in the one-loop RGEs.
As we show below, the low-energy Hamiltonian is then the sum of $N/2$ [$(N-1)/2$ for $N$ odd]
two-leg ladder Hamiltonians corresponding to the band pairs $(j,\bar{\jmath})$ (plus the
Hamiltonian of a single chain for $N$ odd).

\subsubsection{Three-leg ladder}
It is instructive to start with the RG flow of the three-leg ladder in the limit
$v_{1}=v_{3}\ll v_{2}$ (i.e., $t_{\perp}/t\rightarrow\sqrt{2}$), where we can
neglect in the RGEs all contributions, where contractions over the band~2 take
place (see Appendix~A). We then obtain the following results. First, the
renormalization of the two- and single-band interactions of the bands~1 and~3 is
exactly the same as for a two-leg ladder, where the couplings flow towards
universal ratios (for a comparison, see Ref.~\cite{bib:LBFso8})
\begin{eqnarray}
  0<4g_{13}&=&4c_{13}^{\rho}=4f_{13}^{\rho}=c_{13}^{\sigma}=-c_{11}^{\sigma}
    =-c_{33}^{\sigma}
  \nonumber\\
  &=&4u_{1331}^{\rho}=8u_{1133}^{\rho}=u_{1331}^{\sigma},
  \label{eq:TwoLegRatios}
\end{eqnarray}
and
\begin{equation}
  c_{11}^{\rho}=c_{33}^{\rho}\approx f_{13}^{\sigma}\approx 0,
\end{equation}
and become of the order of the bandwidth $t$ at the temperature scale
$T_{1}\sim te^{-\alpha_{1}v_{1}/U}$ ($\alpha_{1}$ is a constant of the order of 1). Second, the
interactions between the bands 1 and 2 (and 2 and 3) are either not renormalized or flow
to 0. Finally, the scaling of the remaining three-band interactions can be calculated as follows
(note that $u_{2213}^{\rho}$ stays always small).
Defining ${\mathbf v}=(c_{1223}^{\rho},c_{1223}^{\sigma},u_{1223}^{\rho},u_{1223}^{\sigma})$ and
\begin{equation}
 M=\left(\begin{array}{cccc}
      2 & 0.75 &  3 & -0.75  \\
      4 &    0 &  4 & -1     \\
      3 & 0.75 &  2 & -0.75  \\
     -4 &   -1 & -4 &  0
  \end{array}\right),
\end{equation}
we rewrite the RGEs as (the scaling variable $l$ is related to the energy by $T\sim te^{-\pi l}$,
see Appendix~A)
\begin{equation}
  (v_{1}+v_{3})\frac{d{\mathbf v}}{dl}=g_{13}(l)M{\mathbf v},
\end{equation}
where, using Eq. (\ref{eq:TwoLegRatios}), we have expressed the interactions of band~1 and~3 
in terms of $g_{13}$. The eigenvalues of $M$ are $(7,-1,-1,-1)$ and
\begin{equation}
  \frac{1}{v_{1}+v_{3}}\int_{l_{0}}^{l}g_{13}(l^{\prime})dl^{\prime}
   =\frac{1}{12}\ln\left[\frac{g_{13}(l)}{g_{13}(l_{0})}\right],
\end{equation}
where $l_{0}<l$ denotes the scale at which the interactions of the bands 1 and 3 are sufficiently
close to the two-leg ladder ratios. The three-band interactions therefore stay in fixed ratios and
\begin{equation}
  \frac{c_{1223}^{\rho}(l)}{g_{13}(l)}
   =\frac{c_{1223}^{\rho}(l_{0})}{g_{13}(l_{0})}
   \left[\frac{g_{13}(l_{0})}{g_{13}(l)}\right]^{5/12}
   \propto\left[\frac{U}{g_{13}(l)}\right]^{5/12}
\end{equation}
such that the three-band interactions are arbitrary small at the scale where $g_{13}\sim t$.

\begin{figure}[t]
  \centerline{
    \psfig{file=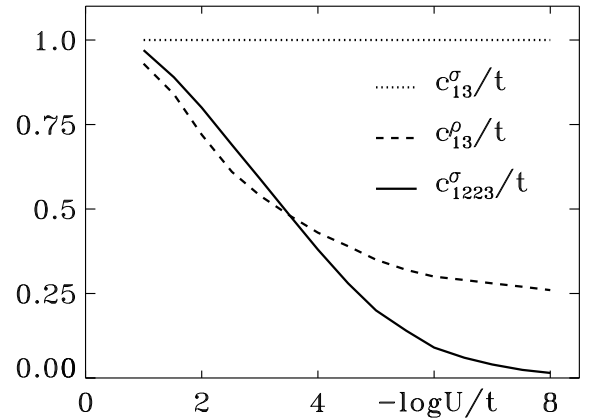,width=8cm,height=6cm}}
  \vspace{1mm}
  \caption{The figure shows (for $t=t_{\perp}$) the strong coupling value as a function
  of the initial value $U/t$, where we fix the strong coupling value of $c_{13}^{\sigma}$
  at the bandwidth $t$. The three-band coupling $c_{1223}^{\sigma}$ then
  decreases to 0 (the flow of the other three-band couplings is similar), as the
  initial value is reduced, and the coupling $c_{13}^{\rho}$ approaches its asymptotic value,
  $c_{13}^{\rho}=c_{13}^{\sigma}/4$. We note that also for the half-filled two-leg
  ladder~[11], the flow to the asymptotic ratios is very slow, i.e., one
  has to take very small initial values~$U/t$.
  }
  \label{f:Flow}
\end{figure}

For other ratios of the velocities $v_{1}/v_{2}<1$, we have performed a
numerical integration of the RGEs: Plotting the strong coupling value
as a function of the initial value $U/t$ (we fix the strong coupling value
of $c_{13}^{\sigma}$ at the bandwidth $t$),
we find that the couplings given in Eq.~(\ref{eq:TwoLegRatios})
always grow up to the bandwidth $t$, approaching their universal
ratios, while all the other couplings stay an order of magnitude smaller, i.e.,
the asymptotic behavior of the couplings for $U/t\rightarrow 0$ is the same
as for $v_{1}\ll v_{2}$ (see Fig.~\ref{f:Flow}). We note that in
Ref.~\cite{bib:Arrigoni} a flow to strong coupling of the three-band
interactions has been found for $t_{\perp}/t<0.86$; we only obtain that for
$v_{1}\approx v_{2}$, i.e., in the strongly anisotropic limit
$t_{\perp}/t<0.2$~\cite{bib:ArComm}.

To conclude our RG analysis of the three-leg ladder: The couplings of the
bands~1 and~3 scale towards the two-leg ladder ratios, become of the order of the
bandwidth~$t$ at the temperature scale $T_{1}\sim te^{-\alpha_{1}v_{1}/U}$, and at (and
below) $T_{1}$, the bands 1 and 3 are \emph{decoupled} from the band~2.
The Hamiltonian becomes therefore $H=H_{13}+H_{s}$. The first
term $H_{13}$ is the Hamiltonian of a two-leg ladder (see Ref.~\cite{bib:LBFso8}) and
the second term $H_{s}$ is the Hamiltonian of a single ``chain''.
At the scale $T_{1}$, the couplings of $H_{13}$ are of the order of $t$ and
all the charge and spin-modes of the bands~1 and~3 acquire
a gap (this result is obtained by bosonization, see 
Ref.~\cite{bib:LBFso8}). The couplings of $H_{s}$
are still small at $T_{1}$, i.e., $0<c_{22}^{\rho},u_{22}^{\rho}\ll t$,
such that the charge and spin-modes of band~2 remain gapless.

For a (small but) \emph{finite} $U$, we then have to investigate the system at energies
\emph{below} $T_{1}$. For the Hamiltonian $H_{s}$ it is well-known that it
has a charge gap below a temperature $T_{2}\sim t e^{-\alpha_{2}v_{2}/U}$
(since $\alpha_{2}\approx\alpha_{1}$ we write in the following $\alpha$ for
both of them) and that the spin-excitations remain gapless
\cite{bib:SchulzRev}. Since the two-leg part has no gapless mode, the final
phase is C0S1 --- the same as for the strongly interacting limit, i.e., a
three-leg spin-ladder.

\subsubsection{Four-leg ladder}
For the four-leg ladder, we find a similar behavior. The couplings of the bands~1 and~4 grow,
become of the order of the bandwidth $t$ at the energy scale $T_{1}\sim te^{-\alpha v_{1}/U}$,
and stay in the same ratio as for a two-leg ladder, while all the other couplings remain small.
In the limit $v_{1}=v_{4}\ll v_{2}=v_{3}$, the RGEs of the four-band couplings become the same
as for the three-band couplings and consequently, similarly as above, the four-band interactions
stay in fixed ratios and
\begin{equation}
  c_{1234}^{\rho}/c_{14}^{\rho}\propto \left(U/c_{14}^{\rho}\right)^{5/12}.
\end{equation}
At the temperature scale $T_{1}$, the bands~1 and~4 are therefore decoupled from the bands~2 and~3.
Again, for a finite $U$, we then have to study the flow of the couplings of band 2 and 3 at
energies \emph{below} $T_{1}$. We find that they flow at the scale
$T_{2}\sim te^{-\alpha v_{2}/U}$ also to a two-leg ladder fixed point (the flow to the two-leg
ladder fixed point takes place for a wide range of initial values, also for non-Hubbard-like). The
(bosonized) Hamiltonian is then the sum of two two-leg ladder Hamiltonians,
$H=H_{14}+H_{23}$. All charge and spin excitations are therefore gapped (as it
is the case in the Heisenberg four-leg spin ladder).

\subsubsection{The case $N>4$}
Analyzing the RGEs for $N>4$, we conjecture that this successive freezing out of band pairs
$(j,\bar{\jmath})$ (plus a single band for $N$ odd) at the characteristic energies
$T_{j}\sim te^{-\alpha v_{j}/U}$ takes place also for (small) $N>4$ with $U\ll t/N$. Again,
the $T_{j}$ are a result of the one-loop RGEs: the couplings of the band pair
$(j,\bar{\jmath})$ scale towards a two-leg ladder fixed point and become of the order of the
bandwidth $t$ at $T_{j}$. Note that we have a hierarchy of energy-scales
\begin{equation}
  T_{1}>T_{2}>\ldots>T_{r},
  \label{eq:HES}
\end{equation}
where for $N$ even $r=N/2$ and for $N$ odd $r=(N+1)/2$.

To conclude this section: At low energies the Hubbard-ladder Hamiltonian becomes for $N$ even
the sum of $N/2$ two-leg ladder Hamiltonians and for $N$ odd the sum of $(N-1)/2$ two-leg
ladder Hamiltonians plus the Hamiltonian of a single chain
\begin{equation}
  H=\sum_{j}H_{j\bar{\jmath}}+\delta_{N,{\mathrm odd}}H_{s},
  \label{eq:BosNL}
\end{equation}
where for $N$ odd, the single chain [the band $(N+1)/2$] has the lowest energy-scale and for $N$
even the two-leg ladder Hamiltonian $H_{N/2,N/2+1}$ [corresponding to the band pair $(N/2,N/2+1)$].
The (half-filled) two-leg ladder Hamiltonians have no gapless excitations and the single-chain
Hamiltonian present for odd $N$ has only one gapless spin-mode. The phases at half-filling are thus
the same as for the Heisenberg spin-ladders~\cite{bib:DagRice}. Since the Hubbard model converges
for large $U$ onto the $t$-$J$ model (which is at half-filling equivalent to the Heisenberg
model)~\cite{bib:GJR}, the phases of the half-filled $N$-leg Hubbard ladders are the same for small
and large $U$.

We would like to emphasize that we have obtained the final result, Eq.~(\ref{eq:BosNL}),
describing \emph{decoupled band pairs}, from an explicit analysis of the \emph{coupled}
RGEs of the \emph{interacting} $N$-band problem.

\section{Hole doped ladders}
Since the low-energy cutoff is typically of the order of $te^{-t/U}$, the effect of the umklapp
interactions has to be taken into account for low dopings, $\delta<e^{-t/U}$. It is then
advantageous to take the \emph{half-filled} ladder as a starting point to study the
effect of doping. Similarly as done in Refs.~\cite{bib:KLLS,bib:FRev}, we model the doping by
adding a chemical potential term $-\mu Q$, where $Q$ is the total charge.

In a single chain, a hole is decomposed in two quasi-particles, one with charge but no
spin (holon, here the fermionic ``kink'') and the other with spin but no charge (spinon).
The quasi-particles of a two-leg ladder are bound hole-pairs (singlet of charge 2).
At low doping, these hole-pairs behave as hard-core bosons, which are equivalent to spinless
fermions. In Sec. III A, we show that the lightly doped $N$-leg ladders can indeed be
described by an effective $N/2$-band [$(N+1)/2$-band for $N$ odd] model of spinless fermions.
We then study the case when only one of these bands is doped. In Sec. III B, we increase the
hole doping such that many bands become subsequently doped.

\subsection{One band (pair) doped}
For the following discussion, we use the \emph{bosonized} form of the Hamiltonian
(\ref{eq:BosNL}). Bosonization~\cite{bib:Haldane,bib:SchulzRev} is the method of
rewriting fermionic creation (annihilation) operators in terms of field operators
$\Phi_{\alpha}$ and $\Pi_{\alpha}$ satisfying the commutation relation
$[\Phi_{\alpha}(x),\Pi_{\alpha}(y)]=i\delta(x-y)$ ($\alpha$ labels the charge and
spin modes for the different bands; for the application of
bosonization to ladders in the limit $U\ll t_{\perp},t$, see, e.g.,
Refs.~\cite{bib:LBFso8,bib:LBF,bib:FRev}).
It is convenient to introduce the dual field of $\Phi_{\alpha}$,
$\partial_{x}\theta_{\alpha}=\Pi_{\alpha}$. For fermions on a chain or ladder,
bosonization applies in the continuum limit.

It is straightforward to
generalize the effective Hamiltonian for the half-filled two-leg ladder obtained in
Ref.~\cite{bib:FRev} to $N/2$ [$(N-1)/2$ for $N$ odd] two-leg ladder Hamiltonians.
We introduce the fields
\begin{equation}
  \Phi_{+j}=\frac{1}{2}\left(\Phi_{\rho j}+\Phi_{\rho\bar{\jmath}}\right)\;\;
  {\mathrm and}\;\;\Pi_{+j}=\Pi_{\rho j}+\Pi_{\rho \bar{\jmath}}
\end{equation}
combining the charge fields of the band pair $(j,\bar{\jmath})$. Dropping bare energy
terms, the Hamiltonian then reads
\begin{eqnarray}
  H&=&\sum_{j}\int dx\biggl\{\frac{u_{+j}}{2}
    \left[\frac{2}{K_{+j}}(\partial_{x}\Phi_{+j})^{2}+\frac{K_{+j}}{2}\Pi_{+j}^{2}\right]\biggr.
  \nonumber\\
  &&\biggl.-g_{+j}\cos(\sqrt{8\pi}\Phi_{+j})-2\mu\partial_{x}\Phi_{+j}\biggr\},
  \label{eq:HamSG}
\end{eqnarray}
plus for odd $N$ the Hamiltonian of a single chain (dropping the gapless spin-part)
\begin{eqnarray}
  \int dx\biggl\{\frac{u_{\rho s}}{2}
    \left[\frac{1}{K_{\rho s}}(\partial_{x}\Phi_{\rho s})^{2}+K_{\rho s}\Pi_{\rho s}^{2}\right]\biggr.
  \nonumber\\
    \biggl.-g_{\rho s}\cos(\sqrt{8\pi}\Phi_{\rho s})-\mu\partial_{x}\Phi_{\rho s}\biggr\}.
  \label{eq:HamSGNodd}
\end{eqnarray}
The coupling $g_{+j}$ is of the order of $g_{j\bar{\jmath}}$, the $K_{\alpha}$ are the Luttinger
liquid parameters (at half-filling,  $K_{+j}\rightarrow 1$ and
$K_{\rho s}\rightarrow 1/2$ \cite{bib:Schulzxxz,bib:SchulzRev}), and the $u_{\alpha}$ are
the velocities of the charge modes. The charge (doping) is given by
\begin{equation}
  Q=\sqrt{\frac{2}{\pi}}\int dx\sum_{j}2\partial_{x}\Phi_{+j}+\delta_{N,{\mathrm odd}}\partial_{x}\Phi_{\rho s}.
\end{equation}
We note that (\ref{eq:HamSG}) and (\ref{eq:HamSGNodd}) are the usual Hamiltonians describing a
commensurate-incommensurate
transition~\cite{bib:MIT}. The fields of the band pair $(j,\bar{\jmath})$, $\Phi_{+j}$ (and of the
single chain $\Phi_{\rho s}$), become pinned at the energy-scales
$T_{j}\sim te^{-\alpha v_{j}/U}$. We therefore define effective gaps for the bands
by $\Delta_{j}\sim T_{j}$, resulting in a hierarchy of gaps (see Eq.~(\ref{eq:HES}))
\begin{equation}
  \Delta_{1}>\Delta_{2}>\ldots>\Delta_{r}.
\end{equation}

\begin{figure}[b]
  \centerline{
    \psfig{file=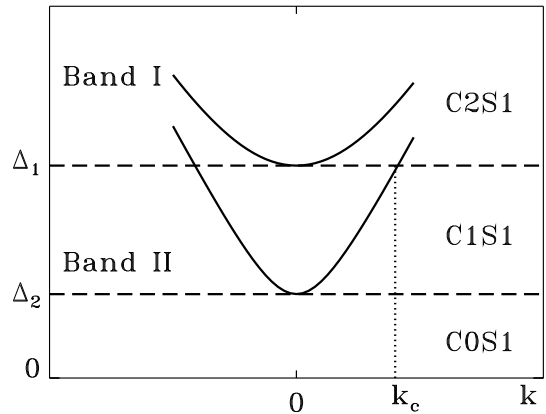,width=8cm,height=6cm}}
  \vspace{2mm}
  \caption{The lightly doped three-leg ladder can be described by a two-band model of
   spinless fermions, where the lower band~II contains the charge part of holes and
   the upper band~I paired holes (singlets). At half-filling the phase is C0S1; upon
   doping, the charge part of the holes enters the band~II and the phase is C1S1. For
   dopings $\delta>\delta_{c}$ (here, $\pi\delta_{c}=|k_{c}|$), paired holes enter also
   the band~I and the phase becomes C2S1.}
  \label{f:TwoBands}
\end{figure}

The equations of motion following from the Hamiltonians~(\ref{eq:HamSG}) and (\ref{eq:HamSGNodd}) are
a set of sine-Gordon (SG) equations. Here, the physical solutions of the SG equations are the
soliton-solutions. Solitons are \emph{fermionic} particles (kinks). The kinks of the single-chain
field $\Phi_{\rho s}$ have a mass $M_{s}\sim\Delta_{s}$ and represent the charge part of single holes,
while the kinks of the two-leg ladder fields $\Phi_{+j}$ have a mass $M_{j}\sim\Delta_{j}$ and
represent paired holes (singlets). The band structure of these solitons reads
\begin{equation}
  \epsilon_{j}(k)=\sqrt{M_{j}^{2}+(v_{j}k)^{2}}.
\end{equation}
At low doping $\delta<e^{-t/U}$, the half-filled $N$-leg ladders are therefore described by
an effective $N/2$-band [$(N+1)/2$-band] model of spinless fermions, where the bottom of
the bands are at the energy-scales $\Delta_{j}$ (for $N=3$, see Fig.~\ref{f:TwoBands}).

At half-filling, $\mu=0$ and minimizing the energy gives $\Phi_{+j}\approx\Phi_{\rho s}\approx 0$.
When doping the first hole (pair), the chemical potential $\mu=\mu(\delta)$ jumps from 0 to
$\Delta_{r}$ and a soliton-solution with charge $Q>0$ minimizes the energy, i.e., the lowest-lying
effective band of spinless fermions becomes doped. For odd $N$, the lowest-lying band is the
single-chain band $(N+1)/2$ and the phase upon doping is a dilute gas of spinons and kinks (C1S1),
while for even $N$, the effective band $N/2$, corresponding to the two-leg-ladder band-pair
$(N/2,N/2+1)$, is doped first and the phase consists of singlets of paired holes (C1S0). Similarly
as for the two-leg ladder, the (singlet) pair-field operator
\begin{eqnarray}
  P_{j}&\propto&\Psi_{Rj\uparrow}\Psi_{Lj\downarrow}+\Psi_{Lj\uparrow}\Psi_{Rj\downarrow}
  \nonumber\\
    &\propto&\exp\left(-i\sqrt{2\pi}\theta_{\rho j}\right)\cos(\sqrt{2\pi}\Phi_{\sigma j})
\end{eqnarray}
has a different sign in band $j$ and $\bar{\jmath}$~\cite{bib:LBFso8},
\begin{equation}
  P_{j}P_{\bar{\jmath}}^{\dagger}<0,
\end{equation}
such that the C1S0 phase has a $d$-wave like symmetry.

It is instructive to rewrite the band operators $\Psi_{hjs}$ respectively the pair-field operators
$P_{j}$ in terms of the chain operators $d_{hjs}$. For $N=3$,
\begin{equation}
  \Psi_{h2s}=\frac{1}{\sqrt{2}}(d_{h1s}-d_{h3s})
\end{equation}
and we find that the holes are situated on the outer legs. Both the phase and the location of the
holes is in agreement with previous numerical treatments~\cite{bib:ThreeLegLD,bib:WSThreeLeg}.
For $N=4$, using $P_{2}P_{3}^{\dagger}\approx -1$, we obtain
\begin{eqnarray}
  P_{2}&\propto&0.22\left(d_{R1\uparrow}d_{L2\downarrow}
    +d_{R2\uparrow}d_{L1\downarrow}+R\leftrightarrow L\right)
  \nonumber\\
    &&+0.22\left(d_{R3\uparrow}d_{L4\downarrow}+d_{R4\uparrow}d_{L3\downarrow}+R\leftrightarrow L\right)
  \nonumber\\
    &&-0.36\left(d_{R1\uparrow}d_{L4\downarrow}+d_{R4\uparrow}d_{L1\downarrow}+R\leftrightarrow L\right)
  \nonumber\\
    &&-0.14\left(d_{R2\uparrow}d_{L3\downarrow}+d_{R3\uparrow}d_{L2\downarrow}+R\leftrightarrow L\right),
\end{eqnarray}
such that the singlets are on the top two legs, the bottom two legs, on the legs 1 and 4, and with
lowest probability on the legs 2 and 3, similarly as found in Ref.~\cite{bib:WStJ} for the $t$-$J$
model. Finally, also the phases for $N=5,6$ are in agreement with numerical works~\cite{bib:WStJ}.
Since the $t$-$J$ model is the large $U$ limit of the Hubbard
model~\cite{bib:GJR}, we conclude that at and close to half-filling the phases of the $N$-leg
Hubbard ladders are the same for small and large $U$. For the two-leg ladder, such a ``universal''
behavior has already been noted in Ref.~\cite{bib:Schulzxxz}.

\subsection{Many bands doped}
When increasing doping, the chemical potential $\mu=\mu(\delta)$ increases too and the hole-pairs
enter subsequently also the higher-lying (effective) bands of spinless fermions. We thus
obtain a series of critical dopings $\delta_{cj}$, where the first hole-pair enters the band $j$.
For $N=3,4$ $\delta_{c1}$ is estimated as
\begin{equation}
  \delta_{c1}\sim\sqrt{e^{-2\alpha v_{1}/U}-e^{-2\alpha v_{2}/U}}\approx e^{-\alpha v_{1}/U}.
\end{equation}
Since for $\delta=\delta_{cj}$, $\partial\mu/\partial\delta=0$, the compressibility
\begin{equation}
  \kappa\propto\frac{\partial\delta}{\partial\mu}
\end{equation}
diverges. The phase transitions at $\delta_{cj}$ belong to the same universality class as
the commensurate-incommensurate transition~\cite{bib:MIT}. We note that our qualitative
estimate for $\delta_{cj}$ does not allow for a comparison with numerical works.

The effective band $j$ corresponds to the band pair $(j,\bar{\jmath})$, such that for a given doping,
the band pairs $(N/2,N/2+1),\ldots,(j,\bar{\jmath})$ are conducting, while the band pairs
$(j-1,\bar{\jmath}+1),\ldots,(1,N)$ form still an ISL. Next, we interpret this result by mapping
each band (pair) on a 2D FS, see Figs.~\ref{f:FermiS3Leg} and~\ref{f:FermiS4Leg}. Using the
dispersion relation (\ref{eq:DispRelN}), the longitudinal Fermi momentum of the band $j$ is
given by
\begin{equation}
  k_{Fj}=\pi-\arccos\left[\frac{t_{\perp}}{t}\cos\left(\frac{\pi j}{N+1}\right)\right]
\end{equation}
and the corresponding transverse Fermi momentum reads
\begin{equation}
  k_{Fyj}=\frac{\pi j}{N+1}.
\end{equation}
Upon doping, the holes enter (for $t_{\perp}\sim t$) then first near the wave vector $(\pi/2,\pi/2)$
and finally for increasing doping also near $(\pi,0)$ and $(0,\pi)$. The FS thus becomes
truncated, similarly as in the 2D case studied in Ref.~\cite{bib:FRS}.

For increasing doping, a crossover takes place to the situation away from half-filling, where
the umklapp processes can be neglected. It is instructive to study the phases for $N=3,4$, when
both effective bands of spinless fermions are doped and to compare with the phase away from
half-filling. In the case $N=3$ (see Fig.~\ref{f:TwoBands}), the \emph{different charges} in the effective
bands~I and~II \emph{forbid} scattering processes of the form
\begin{equation}
  \Psi_{RI}^{\dagger}\Psi_{RII}\Psi_{LI}^{\dagger}\Psi_{LII}+I\leftrightarrow II,
  \label{eq:TwoBdScatt}
\end{equation}
where $\Psi_{R/Lb}^{\dagger}$ creates a spinless fermion in band $b=I/II$,
since they break U(1) invariance and only the interactions
\begin{equation}
  \Psi_{RI}^{\dagger}\Psi_{RI}\Psi_{LII}^{\dagger}\Psi_{LII}+I\leftrightarrow II
  \label{eq:DensInt}
\end{equation}
are allowed. For weak interactions, we can bosonize Eq. (\ref{eq:DensInt}) resulting in a
density term of the form $\partial_{x}\Phi_{I}\partial_{x}\Phi_{II}$ (plus the same for
the dual fields $\theta_{I,II}$). This term does not reduce the number of gapless modes such
that the phase for $\delta>\delta_{c}$ is C2S1. The same phase has been found by the RG and
bosonization methods away from half-filling~\cite{bib:Arrigoni,bib:LBF}. This contrasts
numerical calculations for the strongly interacting $t$-$J$ model~\cite{bib:WSThreeLeg},
where a spin-gapped phase has been found away from half-filling. Such a phase maybe
appears also for weak interactions, if the condensation to the low-energy phase takes place
for all bands at the same energy-scale (for another idea, see Ref.~\cite{bib:EKZ}).

For $N=4$, we have two types of paired holes with the same charge in the ladder, localized
along the rungs. In contrast to the three-leg case, scattering processes of the form
(\ref{eq:TwoBdScatt}) are allowed. We have shown in Ref.~\cite{bib:ULKLH}, that  in such a
two-band model of spinless fermions with repulsive interactions, slightly away from the
band edge, binding takes place and the C2S0 phase close to the band edge becomes a
4-hole C1S0 phase.
Away from half-filling, by RG and bosonization, the phase C3S2 has been found~\cite{bib:LBF}.
Similarly as already noted in Ref.~\cite{bib:EKZ}, we argue that the C3S2 phase is a result
of the $U\rightarrow 0$ limit considered in Ref.~\cite{bib:LBF}; and indeed, our phases agree
with a numerical treatment of the $t$-$J$ four-leg ladder~\cite{bib:WSFourLeg}. At low doping,
the authors obtained a phase of paired holes and at higher doping levels, four-hole clusters
(or two-pair clusters). We therefore argue that for increasing doping the phase remains
$U$-independent in even-leg ladders, but not in odd-leg ladders.

We finally note that including next nearest neighbor hopping terms etc., the Fermi momenta
of band pairs add up no more exactly to $\pi$, i.e., $v_{j}\approx v_{\bar{\jmath}}$. However,
since the phases are controlled by the relative size of gaps, only a \emph{sufficiently
large perturbation} can qualitatively change the corresponding physics.

\section{Conclusions}
We have studied the doping away from half-filling in $N$-leg Hubbard ladders in the case of a small
on-site repulsion $U$, where the ladders are equivalent to a $N$-band model. At half-filling, the
$N$ Fermi velocities are such that $v_{1}=v_{N}<v_{2}=v_{N-1}<\ldots$. Using RG and bosonization
techniques, we have obtained a hierarchy of energy-scales $\Delta_{j}\sim te^{-\alpha v_{j}/U}$, where
the couplings of the band pair $(j,\bar{\jmath})$ (and of the single band $(N+1)/2$ for $N$ odd) are of
the order of the bandwidth $t$, decoupled from the other bands, and at a two-leg ladder fixed point.
The low-energy Hamiltonian is then the sum of $N/2$ [$(N-1)/2$ for $N$ odd] two-leg ladder Hamiltonians,
corresponding to the band pairs $(j,\bar{\jmath})$, without gapless excitations (plus for odd $N$ the
Hamiltonian of a single chain with one gapless spin-mode). We thus obtain the same phases as for the
Heisenberg spin-ladders.

Upon doping, the holes enter first the band (pair) with the smallest gap $\Delta_{j}$. In odd-leg
ladders, this is the nonbonding band $(N+1)/2$ and the phase is a dilute gas of holons (here, the
fermionic kinks) and spinons (C1S1), while in even-leg ladders, the band pair $(N/2,N/2+1)$ is doped
first and the phase consists of singlets of paired holes (C1S0), reflecting the odd-even effect at
half-filling (i.e., of Heisenberg spin-ladders). The same phases have been found in numerical works
for the $t$-$J$ model~\cite{bib:WStJ,bib:ThreeLegLD,bib:WSThreeLeg,bib:WSFourLeg}, i.e., at and
close to half-filling, the phases of the $N$-leg Hubbard ladders are the same for small and large $U$.

For a finite $U$ and at low doping, the ladders can be described by an effective $N/2$-band model
[$(N+1)/2$-band model for $N$ odd] of spinless fermions, where the (effective) band $j$ corresponds
to the band pair $(j,\bar{\jmath})$ and the bottoms of these bands are at the energy-scales $\Delta_{j}$.
Upon doping, the hole-pairs then subsequently enter these bands and the FS is successively opened,
first near the wave vector $(\pi/2,\pi/2)$ and finally also near $(\pi,0)$ and $(0,\pi)$, similarly
as in the 2D case~\cite{bib:FRS}. In addition, we obtain peaks in the compressibility at critical
dopings $\delta_{cj}$, where the first hole-pair enters the band pair $(j,\bar{\jmath})$.

For increasing doping, the $U$-independence of the phase seems to break down, at least for odd-leg
ladders. However, the $U$-independence and the similarity to the 2D case at and close to half-filling 
support the hope that also in 2D, some of the weak-coupling RG results can be extended to
large $U$~\cite{bib:FRS}.

\acknowledgments{We thank S. Haas, C. Honerkamp, and F.C. Zhang for fruitful discussions throughout this work.}

\appendix

\section{Renormalization Group}
The renormalization group (RG) method is a controlled way of subsequently
eliminating (integrating out) high-energy modes in a given
Hamiltonian. While the noninteracting part $H_{0}$ is at a
\emph{RG fixed point}, couplings of the interacting part $H_{\mathrm{Int}}$
may grow or decrease under a RG transformation (i.e., when lowering the energy).
A subsequent (perturbative) elimination of  high-energy modes in
$H_{\mathrm{Int}}$ then results in RGEs, which give the change
of the coupling when lowering the energy, see Ref.~\cite{bib:Shankar}.

Below, we give the RGEs for the couplings of the three-leg ladder.
Fully taking into account the symmetries [i.e., SU(2), U(1), and
$v_{1}=v_{3}$], there are 21 independent RGEs. Using
\emph{operator product expansion} for the products of the various currents,
the derivation is straightforward \cite{bib:OPE}. Similar RGEs are obtained in the
case $N>3$~\cite{bib:ULup}.

For the half-filled three-leg ladder $v_{1}=v_{3}$, such that
$f_{12}^{\rho,\sigma}=f_{23}^{\rho,\sigma}$, $c_{12}^{\rho,\sigma}=c_{23}^{\rho,\sigma}$,
and $c_{11}^{\rho,\sigma}=c_{33}^{\rho,\sigma}$. The initial values of the two and single
band couplings are given by
\begin{eqnarray}
  f_{13}^{\rho}=c_{11}^{\rho}=c_{13}^{\rho}=\frac{3U}{16},\;
    f_{12}^{\rho}=c_{12}^{\rho}=\frac{U}{8},\;c_{22}^{\rho}=\frac{U}{4},
  \nonumber\\
  u_{1331}^{\rho}=\frac{3U}{8},\;u_{1133}^{\rho}=\frac{3U}{32},\;u_{1331}^{\sigma}=0,
    \;u_{22}^{\rho}=\frac{U}{8},
\end{eqnarray}
and $f_{ij}^{\sigma}=4f_{ij}^{\rho}$, $c_{ij}^{\sigma}=4c_{ij}^{\rho}$. The initial
values of the three-band couplings read
\begin{eqnarray}
  c_{1223}^{\rho}=\frac{U}{8},\;\;c_{1223}^{\sigma}=\frac{U}{2},
  \nonumber\\
  u_{1223}^{\rho}=\frac{U}{4},\;\;u_{2213}^{\rho}=\frac{U}{8},
    \;\;u_{1223}^{\sigma}=0.
\end{eqnarray}
The RGEs take the following form, where the energy-scale is related to $l$ by $T\sim te^{-\pi l}$.
Starting with the above initial values the RGEs are then numerically integrated up to the
scale $l_{c}$, where the first coupling is of the order of the bandwidth $t$. For one-loop
RGEs, $l_{c}=\alpha t/U$, where $\alpha\sim 1$ (this can easily be checked by multiplying
the RGEs with $t/U^{2}$). We use the abbreviations
\begin{eqnarray}
  A_{\pm}&=&\pm\left[\left(c_{1223}^{\rho}\right)^{2}+\frac{3}{16}\left(c_{1223}^{\sigma}\right)^{2}\right]
  \nonumber\\
   &&+\left(u_{1223}^{\rho}\right)^{2}+\frac{3}{16}\left(u_{1223}^{\sigma}\right)^{2}
  \nonumber\\
  B_{\pm}&=&\pm 4\left(c_{1223}^{\rho}c_{1223}^{\sigma}-u_{1223}^{\rho}u_{1223}^{\sigma}\right)
  \nonumber\\
   &&-\left(c_{1223}^{\sigma}\right)^{2}-\left(u_{1223}^{\sigma}\right)^{2}
  \nonumber\\
  C_{\pm}^{\rho,\sigma}&=&\frac{c_{13}^{\rho,\sigma}+f_{13}^{\rho,\sigma}}{2v_{1}}
    \pm\frac{2f_{12}^{\rho,\sigma}}{v_{12}}+\frac{c_{22}^{\rho,\sigma}}{2v_{2}}
  \nonumber\\
  D&=&c_{1223}^{\rho}u_{1223}^{\rho}-\frac{3}{16}c_{1223}^{\sigma}u_{1223}^{\sigma},
\end{eqnarray}
where $v_{12}=v_{1}+v_{2}$.

The two-band forward-scattering interactions are renormalized according
\begin{eqnarray}
  \frac{df_{13}^{\rho}}{dl}&=&\frac{1}{2v_{1}}\left[\left(c_{13}^{\rho}\right)^{2}+\frac{3}{16}
                            \left(c_{13}^{\sigma}\right)^{2}+\left(u_{1331}^{\rho}\right)^{2}\right]
          \nonumber\\
                            &&+\frac{1}{2v_{1}}\left[16\left(u_{1133}^{\rho}\right)^{2}+\frac{3}{16}
                            \left(u_{1331}^{\sigma}\right)^{2}\right]
                            +\frac{A_{+}}{2v_{2}}
  \nonumber\\
  \frac{df_{12}^{\rho}}{dl}&=&\frac{1}{v_{12}}\left[\left(c_{12}^{\rho}\right)^{2}
                            +\frac{3}{16}\left(c_{12}^{\sigma}\right)^{2}
                            +4\left(u_{2213}^{\rho}\right)^{2}\right]+\frac{A_{-}}{v_{12}}
  \nonumber\\
  \frac{df_{13}^{\sigma}}{dl}&=&\frac{1}{2v_{1}}\left[2c_{13}^{\rho}c_{13}^{\sigma}
                               -\frac{1}{2}\left(c_{13}^{\sigma}\right)^{2}
                               -\left(f_{13}^{\sigma}\right)^{2}\right]
               \nonumber\\
                               &&+\frac{1}{2v_{1}}\left[2u_{1331}^{\rho}u_{1331}^{\sigma}
                               -\frac{1}{2}\left(u_{1331}^{\sigma}\right)^{2}\right]
                               +\frac{B_{+}}{4v_{2}}
  \nonumber\\                            
  \frac{df_{12}^{\sigma}}{dl}&=&\frac{1}{v_{12}}\left[2c_{12}^{\rho}c_{12}^{\sigma}
                              -\frac{1}{2}\left(c_{12}^{\sigma}\right)^{2}
                              -\left(f_{12}^{\sigma}\right)^{2}\right]
                              +\frac{B_{-}}{2v_{12}},
\end{eqnarray}
the single-band interactions according
\begin{eqnarray}
  \frac{dc_{11}^{\rho}}{dl}&=&-\sum_{k\neq 1}\frac{1}{2 v_{k}}\left[\left(c_{1k}^{\rho}\right)^{2}
                            +\frac{3}{16}\left(c_{1k}^{\sigma}\right)^{2}\right]
              \nonumber\\
                            &&+\frac{1}{2v_{1}}\left[\left(u_{1331}^{\rho}\right)^{2}
                            +\frac{3}{16}\left(u_{1331}^{\sigma}\right)^{2}\right]
  \nonumber\\
  \frac{dc_{22}^{\rho}}{dl}&=&-\sum_{k\neq 2}\frac{1}{2 v_{k}}\left[\left(c_{2k}^{\rho}\right)^{2}
                            +\frac{3}{16}\left(c_{2k}^{\sigma}\right)^{2}\right]
             \nonumber\\
                            &&+\frac{8}{v_{2}}\left(u_{22}^{\rho}\right)^{2}
                            +\frac{A_{+}}{v_{1}}
  \nonumber\\
  \frac{dc_{11}^{\sigma}}{dl}&=&-\sum_{k\neq 1}\frac{1}{2 v_{k}}\left[
                            \frac{1}{2}\left(c_{1k}^{\sigma}\right)^{2}
                            +2c_{1k}^{\rho}c_{1k}^{\sigma}\right]
                            -\frac{1}{2 v_{1}}\left(c_{11}^{\sigma}\right)^{2}
             \nonumber\\
                            &&-\frac{1}{2v_{1}}\left[2u_{1331}^{\rho}u_{1331}^{\sigma}
                            +\frac{1}{2}\left(u_{1331}^{\sigma}\right)^{2}\right]
  \nonumber\\
  \frac{dc_{22}^{\sigma}}{dl}&=&-\sum_{k\neq 2}\frac{1}{2 v_{k}}\left[
                            \frac{1}{2}\left(c_{2k}^{\sigma}\right)^{2}
                            +2c_{2k}^{\rho}c_{2k}^{\sigma}\right]
             \nonumber\\
                            &&-\frac{1}{2 v_{2}}\left(c_{22}^{\sigma}\right)^{2}
                            +\frac{B_{+}}{2v_{1}},
\end{eqnarray}
the two-band back-scattering interactions according
\begin{eqnarray}
  \frac{dc_{13}^{\rho}}{dl}&=&-\sum_{k}\frac{1}{2 v_{k}}\left(c_{1k}^{\rho}c_{k3}^{\rho}
                            +\frac{3}{16}c_{1k}^{\sigma}c_{k3}^{\sigma}\right)
            \nonumber\\
                            &&+\frac{1}{v_{1}}\left(c_{13}^{\rho}f_{13}^{\rho}
                            +\frac{3}{16}c_{13}^{\sigma}f_{13}^{\sigma}
                            +4u_{1133}^{\rho}u_{1331}^{\rho}\right)
                            +\frac{A_{+}}{2 v_{2}}
  \nonumber\\
  \frac{dc_{12}^{\rho}}{dl}&=&-\sum_{k}\frac{1}{2 v_{k}}\left(c_{1k}^{\rho}c_{k2}^{\rho}
                            +\frac{3}{16}c_{1k}^{\sigma}c_{k2}^{\sigma}\right)
           \nonumber\\
                            &&+\frac{2}{v_{12}}\left(c_{12}^{\rho}f_{12}^{\rho}
                            +\frac{3}{16}c_{12}^{\sigma}f_{12}^{\sigma}\right)
                            +\frac{4}{v_{12}}u_{2213}^{\rho}u_{1223}^{\rho}  
  \nonumber\\
  \frac{dc_{13}^{\sigma}}{dl}&=&-\sum_{k}\frac{1}{2 v_{k}}\left(c_{1k}^{\rho}c_{k3}^{\sigma}
                            +c_{1k}^{\sigma}c_{k3}^{\rho}
                            +\frac{1}{2}c_{1k}^{\sigma}c_{k3}^{\sigma}\right)
                            +\frac{B_{+}}{4v_{2}}
             \nonumber\\
                            &&+\frac{1}{v_{1}}\left(c_{13}^{\rho}f_{13}^{\sigma}
                            +c_{13}^{\sigma}f_{13}^{\rho}-\frac{1}{2}c_{13}^{\sigma}f_{13}^{\sigma}
                            +4u_{1133}^{\rho}u_{1331}^{\sigma}\right)
  \nonumber\\
  \frac{dc_{12}^{\sigma}}{dl}&=&-\sum_{k}\frac{1}{2 v_{k}}\left(c_{1k}^{\rho}c_{k2}^{\sigma}
                            +c_{1k}^{\sigma}c_{k2}^{\rho}
                            +\frac{1}{2}c_{1k}^{\sigma}c_{k2}^{\sigma}\right)
             \nonumber\\
                            &&+\frac{2}{v_{12}}\left(c_{12}^{\rho}f_{12}^{\sigma}
                            +c_{12}^{\sigma}f_{12}^{\rho}-\frac{1}{2}c_{12}^{\sigma}f_{12}^{\sigma}\right)
             \nonumber\\
                            &&+\frac{4}{v_{12}}u_{2213}^{\rho}u_{1223}^{\sigma},         
\end{eqnarray}
and for the three-band interactions, we find
\begin{eqnarray}
  \frac{dc_{1223}^{\rho}}{dl}&=&\frac{u_{1331}^{\rho}u_{1223}^{\rho}
                            +4u_{1133}^{\rho}u_{1223}^{\rho}}{2v_{1}}
                            -\frac{3u_{1331}^{\sigma}u_{1223}^{\sigma}}{32v_{1}}
       \nonumber\\          
                            &&+\frac{2u_{22}^{\rho}u_{1223}^{\rho}}{v_{2}}
                            +c_{1223}^{\rho}C_{-}^{\rho}
                            +\frac{3c_{1223}^{\sigma}}{16}C_{-}^{\sigma}
  \nonumber\\
  \frac{dc_{1223}^{\sigma}}{dl}&=&-\frac{u_{1223}^{\sigma}}{2v_{1}}\left(u_{1331}^{\rho}
                            +4u_{1133}^{\rho}-\frac{1}{2}u_{1331}^{\sigma}\right)
       \nonumber\\                       
                            &&+\frac{u_{1331}^{\sigma}u_{1223}^{\rho}}{2v_{1}}
                            -\frac{2u_{22}^{\rho}u_{1223}^{\sigma}}{v_{2}} 
       \nonumber\\
                            &&+c_{1223}^{\sigma}C_{-}^{\rho}+c_{1223}^{\rho}C_{-}^{\sigma}
                            -\frac{c_{1223}^{\sigma}}{2}C_{+}^{\sigma}.
\end{eqnarray}
The two- and single-band umklapp interactions are renormalized according
\begin{eqnarray}
  \frac{du_{1331}^{\rho}}{dl}&=&\frac{u_{1331}^{\rho}}{v_{1}}\left(f_{13}^{\rho}+c_{11}^{\rho}\right)
                             +\frac{3u_{1331}^{\sigma}}{16v_{1}}\left(f_{13}^{\sigma}-c_{11}^{\sigma}\right)
          \nonumber\\
                             &&+\frac{4c_{13}^{\rho}u_{1133}^{\rho}}{v_{1}}+\frac{D}{v_{2}}
  \nonumber\\
  \frac{du_{1133}^{\rho}}{dl}&=&\frac{c_{13}^{\rho}u_{1331}^{\rho}+4f_{13}^{\rho}u_{1133}^{\rho}}{2v_{1}}
                             +\frac{3c_{13}^{\sigma}u_{1331}^{\sigma}}{32v_{1}}
                             +\frac{D}{2v_{2}}
  \nonumber\\
  \frac{du_{1331}^{\sigma}}{dl}&=&\frac{u_{1331}^{\rho}}{v_{1}}\left(f_{13}^{\sigma}
                             -c_{11}^{\sigma}\right)
                             +\frac{u_{1331}^{\sigma}}{v_{1}}\left(f_{13}^{\rho}+c_{11}^{\rho}\right)
          \nonumber\\
                             &&-\frac{u_{1331}^{\sigma}}{2v_{1}}\left(f_{13}^{\sigma}
                             +c_{11}^{\sigma}\right)
                             +\frac{4c_{13}^{\sigma}u_{1133}^{\rho}}{v_{1}}
          \nonumber\\
                             &&+\frac{1}{2v_{2}}\left(c_{1223}^{\sigma}u_{1223}^{\sigma}
                             +2c_{1223}^{\sigma}u_{1223}^{\rho}-2c_{1223}^{\rho}u_{1223}^{\sigma}\right)                 
  \nonumber\\
  \frac{du_{22}^{\rho}}{dl}&=&\frac{2c_{22}^{\rho}u_{22}^{\rho}}{v_{2}}
                             +\frac{D}{v_{1}}
\end{eqnarray}
and the three-band umklapp interactions according
\begin{eqnarray}
  \frac{du_{1223}^{\rho}}{dl}&=&\frac{u_{1331}^{\rho}c_{1223}^{\rho}+4u_{1133}^{\rho}c_{1223}^{\rho}}{2v_{1}}
                             +\frac{3u_{1331}^{\sigma}c_{1223}^{\sigma}}{32v_{1}}
          \nonumber\\
                             &&+\frac{4u_{2213}^{\rho}c_{12}^{\rho}}{v_{12}}
                             +\frac{2u_{22}^{\rho}c_{1223}^{\rho}}{v_{2}}
          \nonumber\\
                             &&+u_{1223}^{\rho}C_{+}^{\rho}
                             -\frac{3u_{1223}^{\sigma}}{16}C_{-}^{\sigma}
  \nonumber\\
  \frac{du_{2213}^{\rho}}{dl}&=&\frac{2u_{1223}^{\rho}c_{12}^{\rho}}{v_{12}}
                             +\frac{3u_{1223}^{\sigma}c_{12}^{\sigma}}{8v_{12}}
                             +\frac{4u_{2213}^{\rho}f_{12}^{\rho}}{v_{12}}
  \nonumber\\
  \frac{du_{1223}^{\sigma}}{dl}&=&-\frac{c_{1223}^{\sigma}}{2v_{1}}\left(u_{1331}^{\rho}
                             +4u_{1133}^{\rho}-\frac{1}{2}u_{1331}^{\sigma}\right)
          \nonumber\\
                             &&-\frac{u_{1331}^{\sigma}c_{1223}^{\rho}}{2v_{1}}
                             +\frac{4u_{2213}^{\rho}c_{12}^{\sigma}}{v_{12}}
                             -\frac{2u_{22}^{\rho}c_{1223}^{\sigma}}{v_{2}}
          \nonumber\\
                             &&-u_{1223}^{\rho}C_{-}^{\sigma}
                             +u_{1223}^{\sigma}C_{+}^{\rho}
                             -\frac{u_{1223}^{\sigma}}{2}C_{+}^{\sigma}.
\end{eqnarray}

\end{multicols}


\begin{thebibliography}{99}
\bibitem{bib:ArComm}The origin of this discrepancy is unclear at present;
the author of Ref.~\cite{bib:Arrigoni} has not published his RGEs.
\end{thebibliography}
\end{document}